\documentstyle[prl,aps,epsf]{revtex}
\begin{document}
\draft

\twocolumn[\hsize\textwidth\columnwidth\hsize\csname %
@twocolumnfalse\endcsname

\title
{Dispersion of a single hole in the $t-J$ model}

\author{ T.K.
Lee$^1$, and C.T. Shih$^2$}
\address{
$^{1}$Dept. of Physics, Virginia Tech, Blacksburg, VA\\
and\\
Inst. of Physics, Academia Sinica, Nankang, Taiwan\\
$^{2}$Dept. of Physics, National Tsing Hua Univ., Hsinchu, Taiwan}

\date{\today}
\maketitle
\begin{abstract}
The dispersion of a single hole in the $t-J$ model obtained by 
the exact result of 32 sites 
and the results obtained by self-consistent Born approximation and the
Green function Monte Carlo method can be simply derived by 
 a mean-field theory with d-RVB and antiferromagnetic order
parameters.
In addition, it offers a simple
explanation for the difference observed between those results.  
The presence of the extended van Hove region at
$(\pi,0)$ is a consequence of the d-RVB pairing independent of the
 antiferromagnetic order. Results including $t'$ and $t''$
are also presented and explained consistently in a similar way.
\end{abstract}
                                                                               
]

Important advances in the study of high temperature
superconductors have been made recently by 
 angle-resolved photoemission (ARPES) experiments \cite
{dessau,gofron}.
Extended van Hove singularity (EVHS)
near the Fermi surface of the superconductors is found. In particular, Wells et
al. \cite {wells} have measured the ARPES for the insulating
$Sr_2CuO_2Cl_2$ and find good agreement with the prediction of the
$t-J$ model about the bandwidth and other features. Although there is 
substantial disagreement about the
position of the energy level near ($\pi$,0).

There were many theoretical studies \cite {yulu1} of the 
properties of a single hole in a
quantum antiferromagnet.
The theoretical data used in comparison with ARPES data are obtained
from the $t-J$ model by using the self-consistent Born approximation (SCBA)
\cite {varma,horsch,liu}
to treat the scattering of a single hole with the spin waves of
the Neel state. The result of SCBA shows an EVHS near $(\pi,0)$. 
The SCBA dispersion of the single hole is also reproduced in the Green
function Monte Carlo (GFMC) approach by Dagotto et al. \cite {dag1}. Based on
these results, Dagotto and collaborators \cite {dag2} have argued that
the EVHS is due to the antiferromagnetism and have proposed the
antiferromagnetic-van Hove theory to explain the mechanism of
superconductivity. 

Recently Leung and Gooding \cite{leung} have solved the 
spectra exactly for a single hole in a 32-site square cluster.
The dispersion relation obtained is very similar to the result of SCBA
except they have not observed the exact symmetry resulted from 
folding the Brillouin zone (BZ) into
half due to the antiferromagnetic long range order. 
It should be noticed that this asymmetry is also observed in 
the GFMC result \cite {dag1} for a
$12\times12$ cluster. 

In this paper we will show that the results obtained by SCBA 
\cite {liu} and GFMC 
\cite {dag1}
as well as exact diagonalization \cite {leung} can be easily reproduced
by a mean-field theory including both 
d-wave resonating valence bond (d-RVB) and antiferromagnetic order
parameters. The presence of the EVHS is associated with the d-RVB pairing.

We have also looked at a more general
 $t-J$ model by including next-nearest-neighbor 
hopping, $t'$, and next-next-nearest-neighbor hopping, $t''$.
It is shown below that the complicated dispersions obtained by including
$t'$ and $t''$ are easily understood in terms of the mean-field theory.
A better agreement with 
 the ARPES of the insulating 
$Sr_2CuO_2Cl_2$ \cite{wells} 
could be obtained by tunning the values
of $t'$ and $t''$
\cite {nazarenko,xiang}, 
 but there is still discrepancy.

Several years ago 
several groups \cite{lee,chen,gl}
have used the variational Monte Carlo 
method to study the phase diagram of the t-J
model at small hole density. The antiferromagnetic phase boundary obtained
agrees fairly well with experiments \cite{lee}.
Near half-filling the best trial function has
antiferromagnetism and d-wave resonating-valence-bond (d-RVB)\cite{anderson}
 coexist.
 But no analytical mean-field theory was discussed.

The two order parameters, Neel order and
singlet d-RVB pairing, can be considered simultaneously in a mean-field
theory of the $t-J$ model, which is given by 
\begin{equation}
H=-t\sum_{<i,j>\sigma} (\tilde{c}^+_{i\sigma}\tilde{c}_{j\sigma}+h.c.)
  + J\sum_{<i,j>}({\bf S}_i\cdot{\bf S}_j-{1\over4} n_in_j),
\end{equation}
where $\langle$$i,j\rangle$ is the nearest neighbor pairs and
$\tilde{c}_{i\sigma}=c_{i\sigma}(1-n_{i,-\sigma})$. 
At half-filling, we shall only consider the 
  ${\bf S}_i\cdot{\bf S}_j$
 term which involves
three mean-field order parameters. The staggered magnetization is
$m_s=\langle S^z_A\rangle =-\langle S^z_B\rangle $, where the lattice is divided into A and B
sublattices. The uniform bond order parameters is $\chi=\langle \sum_{\sigma}
c^+_{i\sigma}c_{j\sigma}\rangle$, and d-RVB is
$\Delta=\langle c_{j\downarrow}c_{i\uparrow}-c_{j\uparrow}c_{i\downarrow}\rangle$
if $i$ and $j$ are nearest neighbor sites in x-direction and $-\Delta$
for y-direction. The mean-field Hamiltonian can be diagonalized and this is done recently
by Inaba et al. \cite {inaba} in the slave-boson formalism. They have
examined the mean-field phase diagram of these order parameters.
We shall adopt a slightly different approach.

Instead of taking into account all three order parameters together
 in the mean-field
Hamiltonian,
 we first consider the
staggered magnetization $m_s$  and uniform bond order $\chi$. They produce upper and lower
spin-density-wave (SDW) bands with dispersions:
$\pm\xi_k=\pm(\epsilon_k^2+(Jm_s)^2)^{1\over2}$ 
where $\epsilon_k=-{3 \over 4}J\chi (cosk_x+cosk_y)$
The states in these bands are then paired. At half filling the mean-field Hamiltonian,
in addition to a constant,
is of the form
\begin{eqnarray}
H_{MF} & = & \sum_{{\bf k},\sigma} -\xi_k a^+_{{\bf k}\sigma}
 a_{{\bf k}\sigma}
 + \sum_{{\bf k},\sigma} \xi_k b^+_{{\bf k}\sigma}
 b_{{\bf k}\sigma} \nonumber \\ 
       & + & \sum_{\bf k} \Delta_k (a_{{\bf -k}\uparrow}
 a_{{\bf k}\downarrow}
 - b_{{\bf -k}\uparrow}
 b_{{\bf k}\downarrow})+h.c.
\end{eqnarray}
where $\Delta_k={3\over 4}J\Delta d_k$, and $d_k= cosk_x-cosk_y$.
The sum is taken over the sublattice BZ (SBZ).
The operators of the lower and upper SDW bands are related to the
original $c$ operators by: 
$a_{{\bf k}\sigma}=\alpha_{\bf k}c_{{\bf
k}\sigma}+\sigma\beta_{\bf k}c_{{\bf k}+{\bf Q}\sigma}$, and
$b_{{\bf k}\sigma}=-\sigma\beta_{\bf k}c_{{\bf
k}\sigma}+\alpha_{\bf k}c_{{\bf k}+{\bf Q}\sigma}$ respectively. We set 
${\bf Q}=(\pi,\pi)$ for the  commensurate SDW
  state, 
$\alpha_{\bf k}^2={1\over 2}(1-{\epsilon_k
\over \xi_k})$,
and $\beta_{\bf k}^2={1\over 2}(1+{\epsilon_k
\over \xi_k})$.

$H_{MF}$ can be diagonalized separately for the lower and upper SDW
bands. For the lower band, in addition to a constant, it becomes 
$ \sum_{\bf k} -E_k (f^+_{1{\bf k}}
 f_{1{\bf k}}-f^+_{2{\bf k}}f_{2{\bf k}})$ where
 $f_{1{\bf k}}=u_k a_{{\bf k}\uparrow}-v_k a^+_{-{\bf k}\downarrow}$,
 $f_{2{\bf k}}=v_k a_{{\bf k}\uparrow}+u_k a^+_{-{\bf k}\downarrow}$,
and
\begin{equation}
 E_k=(\xi_k^2+\Delta_k^2)^{1 \over
2}=(\epsilon_k^2+(Jm_s)^2+\Delta_k^2)^{1 \over 2}. 
\end{equation}
The coherence factors are
$u_k^2={1\over 2}(1+{\xi_k
\over E_k})$,
and $v_k^2={1\over 2}(1+{\xi_k
\over E_k})$. In terms of the operators $a^+_{{\bf k}\sigma}$, the
ground state has the familiar BCS form. Similarly, 
the upper SDW band also forms two bands with identical dispersion 
 $\pm E_k$ as those of $f_{1{\bf k}}$ and $f_{2{\bf k}}$. 
If we had not  chosen zero chemical potential for the half-filled
case, there would be four non-degenerate bands as shown by Inaba et al. \cite
{inaba}. 

In the above approach, the local constraint of no doubly
occupied sites is satisfied for the whole lattice on the average
but it is not for each individual site. To obtain a more
accurate quantitative result for the Hamiltonian $H$, we shall use the
variational Monte
Carlo (VMC) method that satisfies the constraint exactly.
The ground state of $H_{MF}$ is simply given by the
product of two BCS-like  wave functions, one for the lower SDW band and
the other for the upper SDW band. 
 In the presence of the constraint the wave function has the form
\begin{equation}
 |\Psi_0\rangle = Pd(\sum_k (A_k a^+_{{\bf k}\uparrow}a^+_{{\bf -k}\downarrow}
 +B_k b^+_{{\bf k}\uparrow}b^+_{{\bf -k}\downarrow}))^{N_e/2} |0\rangle 
\end{equation}
where $N_e$ is the total number of electrons and coefficients 
$A_k=(E_k+\xi_k)/\Delta_k$ and 
$B_k=-(E_k-\xi_k)/\Delta_k$. The projection operator $Pd$ enforces the
constraint of no doubly occupied sites. 
In this wave function there are two variational
parameters: $\Delta/\chi$ and $m_s/\chi$.
In the absence of staggered order, $m_s$,
this is exactly the same RVB wave function used by Gros \cite {gros}. 
Without pairing
this wave function describes the SDW state.  $|\Psi_0\rangle$ is similar to
the trial wave function used by Chen et al. \cite {chen} but with a slightly
lower energy. It has about $-0.332J$ per bond which is within one percent of
the best estimate of the ground state energy of the Heisenberg model.
The success of $|\Psi_0\rangle$ gives support to the mean-field theory that
derives Eq. (4).

According to the mean-field Hamiltonian $H_{MF}$
discussed above, creating a hole is to take away a quasiparticle
$f_{1{\bf k}}$ from the lower SDW band in the ground state, or the
corresponding one in the upper SDW band. Hence the energy of such a state is 
just $-E_g+E_k$, where $-E_g$ is the ground state energy at
half-filling. 
Interestingly, the quasiparticle
dispersion actually has the similar form as the dispersion of the hole
obtained by SCBA. 
In Fig.1 we show that the energy
dispersion of a single hole obtained by Liu and Manousakis \cite {liu} by using SCBA for a $20\times20$ lattice with $J=0.2$
can be fitted quite well by $E_k-E_0$ with parameters, $\chi=5.47$,
$m_s=13.2$ and $\Delta=2.27$.  $E_0=5.37$ is just a constant shift. 
In this paper, the energy unit is $t$. 
However, the value of $m_s$ is unphysically
large. This is mainly due to our neglect of the constraint in deriving 
Eq. (3). The effct of antiferromagnetism is grossly overestimated.
A more quantitative approach is to use the renormalized mean-field
theory  \cite{zhang} by taking into account the constraint a little bit more
carefully.
$m_s$ will be multiplied by a renormalization factor. Instead of 
pursuing this approach we shall use the variational method
to calculate the dispersion numerically. Consequently, 
as shown below, more physically reasonable
values of parameters are obtained. 
Here, the emphasis is that all the interesting features of the
dispersion obtained by SCBA is quite consistent with the 
 form of $E_k$. 

\begin{figure}[m]
\epsfysize=4.25cm\epsfbox{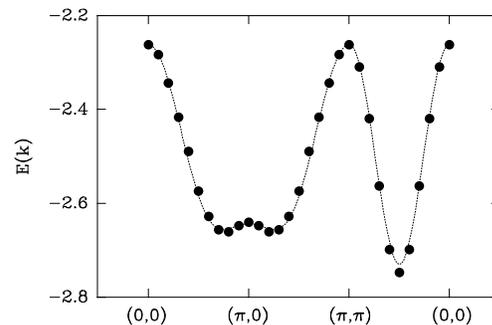}
\caption{
Energy dispersion of a hole in the $t-J$ model 
obtained by Liu and Manousakis [7] 
with SCBA on an
$20\times20$ cluster (solid circles), for $J=0.2$, and $t=1$. 
Dotted line is the fitted result of $E_{k}-E_0$ with $E_0=5.37$,
$\chi=5.47$, $\Delta=2.27$, and $m_s=13.2$.}
\label{ct1p}
\end{figure}

To have
a more accurate description of the single
hole dispersion,
 we shall use the power method 
in addition to the VMC method.
The power method
\cite {chen1}, which is essentially the same as the GFMC method,
 applies many powers of Hamiltonian 
to a trial wave function (TWF) to project
out the ground state with the same symmetry as the TWF. To increase
the convergence the method is modified by combining with the Lanczos
method, we call it the power-Lanczos (PL) method \cite {chen2,heeb}. 

Following the mean-field theory discussed above, we can easily construct
 a variational wave function for a hole with momentum 
$\bf k$ and $S_z=1/2$. This function  
 has $N_e/2-1$ singlet pairs of electrons
 and a single unpaired electron with momentum $\bf k$ and $S_z=1/2$, 
\begin{equation}
|\Psi_1\rangle = Pd~c^+_{{\bf k}\uparrow}
(\sum_q {}' (A_q a^+_{{\bf q}\uparrow}a^+_{{\bf -q}\downarrow}
 +B_q b^+_{{\bf q}\uparrow}b^+_{{\bf -q}\downarrow}))^{{N_e \over2}-1} |0\rangle 
\end{equation}
where the prime on the summation symbol indicates that the momentum $\bf
k$ is excluded from the sum if $\bf k$ is within the SBZ, otherwise, $\bf
k$-$\bf Q$ is excluded. Notice that $|\Psi_1\rangle$ is essentially the same as
$\tilde{c}_{-{\bf k}\downarrow}|\Psi_0\rangle$.

The energies obtained by the VMC method
for an $8\times8$ cluster with $J=0.3$ are shown in Fig. 2 as the open circles.
The variational parameters are $\Delta/\chi=0.3$ and $m_s/\chi=0.056$. 
Notice that the ground state energy for the half-filled lattice ($-22.656J$)
is
subtracted from the data.
 The single-hole state 
has the lowest energy at $(\pi/2,\pi/2)$.
We then apply the PL method to this TWF \cite{note1}
to project it onto the lowest energy state. 
Results are shown as the solid circles in Fig. 2. 
\begin{figure}[m]
\epsfysize=4.25cm\epsfbox{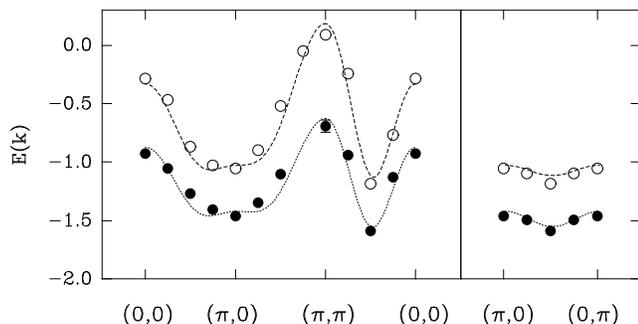}

\caption{Energy dispersion of a hole in the $t-J$ model on an
$8\times8$ cluster, for $J=0.3$, and $t=1$. Open circles
are VMC results by using $|\Psi_1\rangle$, solid circles obtained by
applying the PL method to this wave function. 
$E_{1k}-E_0$ are plotted as the dotted lines. 
For the lower curve 
$\chi=6.92$, $\Delta=2.71$, $m_s=18.84$, $E_0=7.43$, and
$t_{eff}=0.03$} 
\label{ct2p}
\end{figure}

The results represented by the solid circles in Fig. 2 are almost
identical with that of Fig.1. For both, the bandwidth is about $2.2J$,
the energy difference between states at $(\pi/2,\pi/2)$ and $(\pi,0)$
is about $0.43J$, and there is an EVHS around $(\pi,0)$. However, there is
one difference. In the SCBA, starting from the classical Neel state,
the hole only hops between the same sublattice. The hole
energy begins with $t^2/J$. States at ${\bf k}$ and ${\bf k+Q}$ have 
the same energy. 
Our result shows an energy 
difference between $(\pi,\pi)$ and $(0,0)$ and a slight asymmetry between 
energies at ${\bf k}$ and ${\bf k+Q}$.
This is due to the non-vanishing hopping matrix element
between the two sublattices which is related with the uniform bond order
$\chi$. Kane et al. \cite{kane} have pointed out this 
difference between a mean-field theory based on a Neel state or a RVB
state. 

Quantitatively, the difference between Figs.1 and 2 can be 
accounted for by introducing a coherent nearest-neighbor hopping in
addition to $E_k$ of Eq. (3),
 $E_{1k}=E_k-2t_{eff}(cosk_x+cosk_y)$. 
Both the solid and open circles in Fig.2 can be fitted very well by a
functional form $E_{1k}-E_0$, where
 $E_0$ is a
constant.
Although that solid circles represent results
much closer to the ground state 
than the results of TWF (open circles), 
the qualitative feature of the dispersion is
essentially unchanged. 

Much more accurate numerical results are 
 obtained in the exact calculation for a 32-site cluster by Leung
and Gooding \cite{leung,note2} and in GFMC for a $12\times
12$  cluster \cite{dag1}. In Fig.3, the 32-site result for $J=0.3$ is shown as
open squares, solid circles are results from GFMC for $J=0.4$. Both
dispersions can be fitted rather well by $E_{1k}-E_0$ discussed above. 

\begin{figure}[m]
\epsfysize=4.25cm\epsfbox{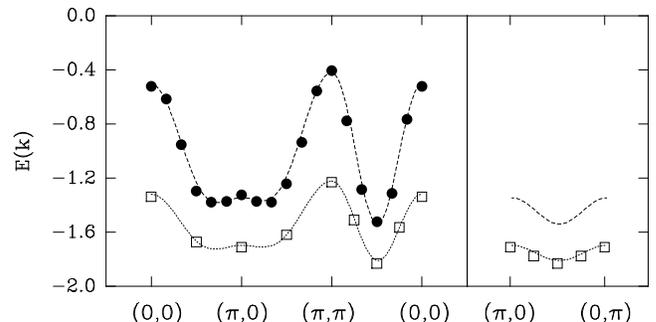}
\caption{
Energy dispersion of a hole in the $t-J$ model on a
32-site cluster (open squares)
 obtained with exact diagonalization [10]
for
 $J=0.3$, and $t=1$. Solid circles are results of GFMC [8]
for a
$12\times12$ cluster with $J=0.4$. Dotted lines are the fitted
$E_{1k}-E_0$ curves. 
For the lower curve (32 sites),
$\chi=5.43$, $\Delta=2.43$, $m_s=17.6$, $E_0=7.09$, and
$t_{eff}=0.012$ 
For the upper curve ($12\times12$ sites),
$\chi=6.87$, $\Delta=2.83$, $m_s=18.4$, $E_0=8.9$, and
$t_{eff}=0.011$.} 
\label{ct3p}
\end{figure}

The results of our trial wave function shown in Fig.2 agree fairly well
with the more accurate results of Fig.3.
This certainly enhances our belief that the mean-field theory or the
trial wave function $|\Psi_1\rangle$ has captured the essential 
physics of the 
one-hole state. 
In particular, we notice that 
 the asymmetrical
  energy dispersion between 
 ${\bf k}$
and ${\bf k}+{\bf Q}$ are present in both 32 and 144 clusters. 
Although a careful and systematic investigation is 
necessary before we can firmly 
establish this asymmetry for larger clusters, this result clearly is
in better agreement with our mean-field theory than the SCBA calculations.
As explained above, even for small clusters, SCBA 
\cite {varma,horsch,liu}
results still have the prefect 
antiferromagnetic symmetry. 

Another interesting result 
is the observation of the flat band region, which is the so called EVHS \cite
{dag2}, 
  near $(\pi,0)$ in Figs.2 and 3. Examining $E_k$ or $E_{1k}$ shows that
EVHS is  
 due to the presence of
$\Delta$ or d-RVB order parameter and it is independent of 
the existence of
long-range antiferromagnetic order. 
The extra features of the
dispersion around $(\pi,0)$ depend on the ratio $\Delta/\chi$.
The d-RVB order parameter has the
largest band gap near $(\pi,0)$. This band gap will produce large
density of states which is reflected by this EVHS.  

In our calculation of the lowest energy state for each wave vector k, it is
not guaranteed that the state we obtained has a finite spectral weight
when a hole is produced from the half-filled ground state. It is therefore 
necessary to examine the spectral weight $Z_{\bf k}$ defined as
$$  Z_{\bf k}=|\langle \Psi_{\bf k}|c_{{\bf k},\sigma}|\Phi\rangle|^2 ,
$$
where $|\Phi\rangle$ is the normalized exact ground state at half filling and
$|\Psi_{\bf k}\rangle$ is the lowest energy state at momentum ${\bf k}$ in the
presence of one hole. Here we use VMC and PL method to calculate
$Z_{\bf k}$. The trial functions are $|\Psi_0\rangle$ of Eq. (4) and 
$|\Psi_1\rangle$ used to obtain Fig. 2.
  The results for 64 sites
are plotted in
 Fig. 4. Open circles are VMC results and solid circles are
results of first order PL \cite{chen2,note1}. We have also included
the exact result of 32 sites \cite {leung} as the open squares. 
For most
wave vectors our results have roughly the right magnitude except near 
${\bf k}=(0,0)$. 
Clearly, we have not yet obtained the exact results for 64 sites and more
higher order power method calculations are needed. 
However, the result is enough
to see the asymmetry between ${\bf k}$ and ${\bf k}+{\bf Q}$,
 which is absent in the SCBA.
 
\begin{figure}[m]
\epsfysize=4.25cm\epsfbox{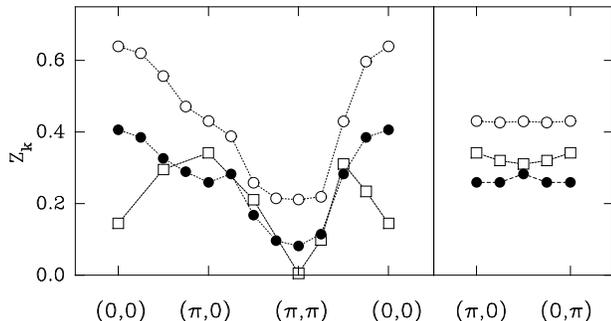}
\caption{Spectral weight $Z_{\bf k}$ as a function of $\bf k$ for the $8\times8$
cluster used in Fig. 2. Open circles are VMC results and solid circles are
by PL1. The exact results 
 of 32 sites are shown as open squares.}
\label{ct4p}
\end{figure}

So far we have assumed that the hole only has the nearest-neighbor 
hopping matrix element. To explain the ARPES \cite{wells}
result, there has been effort \cite {nazarenko,xiang} 
to generalize the model by including the next-nearest-neighbor
hopping, $t'$, and next-next-nearest-neighbor hopping, $t''$.
Fairly complicated results obtained in these cases
can be simply understood in our mean-field theory.
The basic dispersion is determined by the spin interaction. 
The additional contribution by $t$, $t'$ and $t''$ is the same as that 
of the ideal
gas except with a renormalized magnitude.
Hence for the $t-t'-t''-J$ model, the one-hole dispersion 
has the form,
$E_{2k}=E_{1k}-4t'_{eff}cosk_xcosk_y-2t''_{eff}(cos(2k_x)+cos(2k_y))$.
This is verified in Fig. 5.

\begin{figure}[m]
\epsfysize=4.25cm\epsfbox{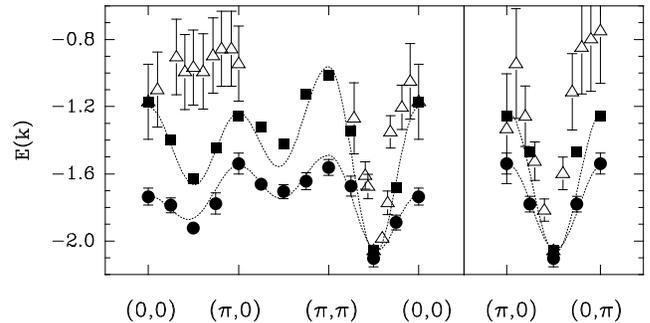}
\caption{Energy dispersion of a hole in the $t-t'-t''-J$ model on an
$8\times8$ cluster, for $J=0.3$, $t'=-0.3$ and $t=1$. 
Circles are for $t''=0$ and squares for $t''=0.2$. Both results are obtained
by using first order PL method on
$|\Psi_1\rangle$[23]. Triangles are the rescaled ARPES data[3].
$E_{2k}-E_0$ are plotted as the dotted lines. 
The parameters are the same as those in Fig. 2, except
$t'_{eff}=0.092$ for the lower curve, and for the upper curve $t'_{eff}=0.065$ 
and $t''_{eff}=-0.055$.}
\label{ct5p}
\end{figure}

In Fig. 5  the first order Lanczos result \cite{note1} of
 using $|\Psi_1\rangle$ for the 
$t-t'-t''-J$ model with $J=0.3$, and
$t'=-0.3, t''=0$ is shown as the circles, $t'=-0.3$ and $t''=0.2$
as the squares.  We have shifted the two sets of data so they match at
${\bf k}=(\pi/2,\pi/2)$.
 The lower dotted curve shows the  fitted
result by $E_{2k}$ with $t'_{eff}=0.092$ and $t''_{eff}=0$. 
 The 
 squares are fitted with $t'_{eff}=0.065$ and $t''_{eff}=-0.055$.
All other fitted parameters are the same as those used in  Fig.2.

Notice that $t'_{eff}$ has the opposite sign of $t'$.
 This can be easily
understood in terms of the d-RVB property of the TWF. When the hole hops
from site $i_x,i_y$ to site $i_x+1,i_y+1$, it rotates a nearest-neighbor
RVB bond in the y-direction to x-direction. Due to  the d-symmetry of the
singlet bond, such a rotation produces a
negative sign in the hopping matrix element. Similarly for $t''$, a nearest-neighbor RVB bond is reversed, and $t''_{eff}$ has the opposite sign.

>From Fig.5 or $E_{2k}$, we can see that the bandwidth, which is the energy
difference between ${\bf k}=(0,0)$ and $(\pi/2,\pi/2)$, would be reduced
if $t'$ is negative and $t''=0$. But for $t''$ to be positive and of 
the similar magnitude as $t'$, the bandwidth 
is about the same as the $t-J$ model.
While for most region of the BZ
 the effect of $t'$ and $t''$ cancels each other,
but near $(\pi,0)$
they add up. Hence the most significant
change between Figs. 2 and 5 is the energy increase for the state
at $(\pi,0)$. This increase is necessary to make the theory
in better agreement with the ARPES data by Wells et al. \cite{wells}. 
The ARPES data are the
  open triangles in Fig. 5. 
We have rescaled the data so that the energies at $(0,0)$
and $(\pi2,\pi/2)$ match that of the squares in Fig. 5. 
Notice that we have not chosen $t'$ or $t''$ in order to match
the experiments. Even if we had used  the matched $t'$ or $t''$, 
the discrepancy
between theory and experiment along $(1,0)$ direction will not be removed.
This can be easily understood by examining $E_{2k}$. The energy at
$(\pi,0)$ may be shifted  by $t'$ and $t''$, but not at $(\pi/2,0)$.
Unlike other well known high
temperature superconductors, 
$Sr_2CuO_2Cl_2$ 
is known to be
fairly difficult to be doped into a superconductor. 
We note that since $(\pi,0)$ is exactly at
the EVHS region with a large density of states, some subtle
difference between materials may become important.
The experimental analysis is further complicated by the recent
observation of the d-wave-like gap structure \cite{loeser,ding} 
near $(\pi,0)$.
More accurate experimental results for 
$Sr_2CuO_2Cl_2$ and other high temperature 
superconductors are needed to clarify
 the discrepancy between theory and experiment.

In summary, by using a mean-field theory that takes into account
both staggered magnetization and d-RVB singlet we have
derived the energy dispersion of a single hole in the $t-J$ and $t-t'-t''-J$
models.  Numerical results obtained from exact diagonalization, GFMC
method and VMC are all in good agreement with the dispersion relation.
In this theory, occurrence of the EVHS near $(\pi,0)$ is due to the 
presence of d-RVB
order parameter. The dispersion does not have the exact sublattice
symmetry observed in the Neel state in which $\bf k$ and $\bf k$+$\bf Q$
are degenerate. 
This particular feature agrees with results obtained by exact 
diagonalization and GFMC method and disagrees with the SCBA.    
 Simple arguments have been provided to account for  this difference.

It should be pointed out that there are other quite successful variational
studies \cite {bm} about the single-hole dispersion. 
Although the wave functions were
constructed in different ways, it was recognized that the spin-flip
terms are  essential in obtaining the right physics. The energy due to
the spin-flip
term is exactly what the RVB order parameters are designed for.
We believe that one of the main reasons why so many different
calculations mentioned in this paper 
all have obtained similar results is that they all have
taken into account the dominant spin-flip effect. The RVB theory
is the easiest way to take into account of this effect right from the
beginning.
It also gives a more intuitively simple interpretation  
of the dispersion and EVHS.

The EVHS has been observed in many HTS. There is not yet a convincing
theoretical reason to explain it. Although the theory presented here is
only valid in the presence of a single hole,  we can a make a few
general comments about the underdoped region. 
In many earlier numerical studies the
d-RVB state has been known to
give a good account of the ground state of the $t-J$ model.
  The d-RVB order parameter would produce a 
 gap and a large density of state  or EVHS
  at $(\pi,0)$. Such a gap may provide a natural explanation for
the spin gap and the gap observed in ARPES \cite{loeser,ding}. 
 Theoretical
work is now in progress to address these issues.

\medskip
We wish to thank Y.C. Chen, P.W. Leung and Yu Lu for many useful discussions. 
Part of the research was conducted using the resources of the Cornell Theory
Center. 
which receives major funding from NSF and NY state with
additional support from ARPA, NIH, IBM and members of the Corporate
Research Institute.
Part of computations were performed at the 
  National Center
for High Performance Computing in Taiwan. We are grateful for their support.
\medskip



\begin{references}
\bibitem{note1} 
 The solid circles shown in Figs. 2 and 5 used the first order
Lanczos to improve the TWF, this is called PL1,  
and then apply two powers of the Hamiltonian. The solid circles in Fig.4 and 
squares in Fig.5 are PL1 without extra powers.
\end{references}
\end{document}